\documentclass[5p,twocolumn]{elsarticle}
\usepackage{amsmath, amssymb, amsfonts}
\usepackage [T1]{fontenc}
\usepackage [utf8]{inputenc}
\usepackage{graphicx}
\usepackage{dcolumn}
\usepackage{bm}
\usepackage{hyperref}
\usepackage{comment}
\usepackage[usenames,dvipsnames]{color}

\newcounter{bla}

\newcommand{\Tc}{\ensuremath{T_\mathrm{c}}}
\newcommand{\e}{\mathrm{e}}

\newcommand{\changes}[1]{#1}

\journal{Computer Physics Communications}

\begin{document}

\title{Microcanonical simulated annealing:\\ Massively parallel Monte Carlo simulations with sporadic random-number generation}

\author[1]{M.~Bernaschi}

\author[2,3]{C.~Chilin}

\author[2]{L.A.~Fernandez}

\author[1]{I.~Gonz\'{a}lez-Adalid Pemart\'{i}n}

\author[3,4]{E.~Marinari}

\author[2]{V.~Martin-Mayor}

\author[5,3,4]{G.~Parisi}

\author[3,4]{F.~Ricci-Tersenghi}

\author[6,7]{J.J.~Ruiz-Lorenzo}

\author[8,9,10]{D.~Yllanes\corref{cor1}}

\address[1]{Istituto per le Applicazioni del Calcolo, CNR, Rome, Italy}

\address[2]{Departamento de Física Teórica, Universidad Complutense, 28040 Madrid, Spain}

\address[3]{Dipartimento di Fisica, Sapienza Università di Roma, 00185 Rome, Italy}

\address[4]{CNR-Nanotec, Rome Unit, and INFN, Sezione di Roma, 00185 Rome, Italy}

\address[5]{International Research Center of Complexity Sciences,
  Hangzhou International Innovation Institute, Beihang University,
  Hangzhou 311115, China}

\address[6]{Departamento de Física, Universidad de Extremadura, 06006 Badajoz, Spain}

\address[7]{Instituto de Computación Científica Avanzada (ICCAEx), Universidad de Extremadura, 06006 Badajoz, Spain}

\address[8]{Fundación ARAID, Diputación General de Aragón, 50018 Zaragoza, Spain}

\address[9]{Instituto de Biocomputación y Física de Sistemas Complejos (BIFI), Universidad de Zaragoza, 50018 Zaragoza, Spain}

\address[10]{Zaragoza Scientific Center for Advanced Modeling (ZCAM), 50018 Zaragoza, Spain}

\date{\today}

\cortext[cor1]{Contact author: david.yllanes@bifi.es}
\begin{abstract}

Numerical simulations of models and theories that describe \changes{complex systems such as spin glasses are becoming increasingly important. Beyond fundamental research, these computational methods also find practical applications in fields like combinatorial optimization.} However, Monte Carlo simulations, an important subcategory of these methods, are plagued by a major drawback: they are extremely greedy for (pseudo) random numbers. The total fraction of computer time dedicated to random-number generation increases as the hardware grows more sophisticated, and can get prohibitive for special-purpose computing platforms.  
We propose here a general-purpose microcanonical simulated annealing (MicSA) formalism that dramatically reduces such a burden. 
The algorithm is fully adapted to a massively parallel computation, as we show in the particularly demanding benchmark of 
the three-dimensional Ising spin glass. We carry out very stringent numerical tests of the new algorithm by comparing our results, obtained on GPUs, with high-precision standard (\emph{i.e.}, random-number-greedy) simulations performed on the Janus II custom-built supercomputer. In those cases where thermal equilibrium is reachable (\emph{i.e.}, in the paramagnetic phase), both simulations reach compatible values. More significantly, barring short-time corrections, a simple time rescaling suffices to map the MicSA off-equilibrium dynamics onto the results obtained with standard simulations.
\end{abstract}

\maketitle


\section{Introduction}

Monte Carlo algorithms have become a ubiquitous tool in science and technology, with an amazing variety of applications ranging from lattice gauge theories in particle physics to high-frequency trading in stock markets. Most of the time, massively parallel computations are needed to achieve the expected goals. A common bottleneck to these parallel computations is in the generation of (pseudo-) random numbers, which has been recognized as critical for large-scale Monte Carlo simulations~\cite{kalos:08}. The difficulties encountered are due to two mutually conflicting goals. On the one hand, a large number of long streams of (sufficiently) independent random numbers are needed for the parallel update of the spins. On the other hand, generating random numbers of sufficiently high quality is computationally intensive (to the point of being a significant limiting factor for numerical simulations).

Modern efforts in high-performance computing have directed a great deal of attention to the numerical simulation of deceptively simple-looking disordered magnetic alloys, known as spin glasses~\cite{mydosh:93,young:98,charbonneau:23,parisi:23,dahlberg:25}. Interest has grown to the point that hardware is being specifically designed to deal with a ``spin-glass Hamiltonian'' through a variety of algorithms and/or physical principles (see, \emph{e.g.}, Refs.~\cite{aramon:19,hayato:19,matsubara:20,takemoto:20,mcgeoch:22,king:23,mcmahon:16,janus:12b,janus:14}).  The choice of this seemingly exotic problem as a framework for more general computations is also connected to the crucial notion of NP-completeness~\cite{papadimitriou:13}, which we briefly recall next. 

The point is here that one of the interesting applications of Monte Carlo methods is to minimize a cost function, for example, by incrementally reducing the ``temperature'' control parameter, starting from a very high value and going down close to zero (where the ``Boltzmann'' measure concentrates at the minimum of the cost function).\changes{This algorithm is the well-known simulated annealing~\cite{kirkpatrick:83}, which can be used to solve, under some conditions, optimization problems in a very effective way~\cite{angelini2019monte,angelini2023limits}}. 

The particular problem at hand is formalized in terms of a cost function that one aims to minimize. For the spin glass, the cost function is the energy of the spin configuration, while, for the traveling salesman, \changes{the cost function is the total travel distance to visit a set of cities}. One wants to find the assignment that minimizes the cost function, often with the help of a computer. If the computational resources (\emph{i.e.}, computer time, memory, etc.) grow with $N$ faster than any polynomial for all known algorithms, the problem is generally regarded as \emph{hard}. A great attention is paid to a small subcategory of these hard problems, termed NP-complete, because if a polynomial scaling algorithm were discovered for any of the NP-complete problems, then a vast family of hard optimization problems would become \emph{easy}. Finding the minimal energy state ---the ground state (GS)--- of an Ising spin-glass Hamiltonian on a non-planar graph is one of such NP-complete problems~\cite{barahona:82b,istrail:00}. 

Understandably, considerable effort has been devoted to the efficient numerical simulations of spin glasses, both on unconventional dedicated hardware (see above) or through the writing of code carefully optimized for its use on GPUs~\cite{yavorskii:12,feng:14,lulli:15b,bernaschi:24,bernaschi:24b} or CPUs~\cite{fernandez:15,billoire:17,billoire:18}. Perhaps unsurprisingly, the generation of random numbers turns out to be a major headache in both computational contexts.  

Here we propose a widely usable microcanonical simulated annealing (MicSA) formalism, which is both extremely frugal in the number of needed random numbers and fully adapted to a massively parallel computation. The method is demonstrated in the rather demanding case of the three-dimensional Ising spin glass. We carry out our simulations on GPUs to demonstrate that our algorithm nicely parallelizes. We start from a microcanonical ensemble for an extended configuration space that comprises the degrees of freedom we are really interested in (the spins) and some auxiliary degrees of freedom traditionally called daemons. We shall term these auxiliary variables \emph{walkers} if they are allowed to wander through the system or \emph{daemons} if each is kept tied to a single lattice site; should the distinction be immaterial, which happens at many points in the discussion, we shall name them daemons/walkers. The reader will find useful analogies between our Monte Carlo formalism and molecular dynamics simulations coupled with a thermostat~\cite{frenkel:23}. 

\begin{figure}
    \centering
    \includegraphics[width=\linewidth]{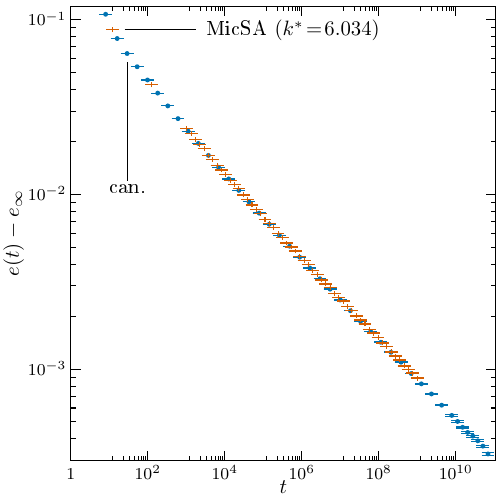}
    \caption{{After a time rescaling, our microcanonical simulated annealing reproduces canonical dynamics}. We study the energy relaxation towards its equilibrium value by plotting the excess energy per spin over its equilibrium value, $e(t)- e_\infty$, versus simulation time $t$, as computed for a spin-glass sample on a $160\times160\times160$ lattice with binary couplings at temperature $T=0.7\approx 0.64 T_\mathrm{c}$ ($T_\mathrm{c}$ is the critical temperature for the phase transition separating the paramagnetic phase at high temperatures from the spin-glass phase at low temperature). The sample is investigated using two dynamics: a standard, random-number greedy Metropolis algorithm (the label \emph{can} in the plot stands for canonical; these data were obtained from the Janus II supercomputer) and the new microcanonical simulated annealing (MicSA) algorithm introduced in this work. Specifically, we are using the version of the algorithm with 6 walkers per spin, see Sec.~\ref{subsect:our-algorithm}. We plot both data sets, subtracting the same estimate for $e_\infty$, which was obtained from the canonical simulation as explained in ~\ref{ap:time-shift}. After the MicSA time is rescaled by the factor $k^*=6.034$  (see Table~\ref{tab:k*}), $t^\text{plot}=k^*\times t^\text{MicSA}$, both energy relaxations are indistinguishable (the naif rescaling would have been $k^\text{naif}=6$, because of the 6 walkers per spin).}
    \label{fig:E}
\end{figure}

A microcanonical framework with just \emph{one} daemon was proposed by Creutz~\cite{creutz:83}, precisely to minimize the need for random numbers. Unfortunately, Creutz's ensemble leads to a rather significant dynamic slowing down in the context of spin glasses~\cite{ruizlorenzo:00}. {To overcome this dynamic slowdown in non-disordered systems, Rummukainen and other authors proposed using a number of daemons proportional to the number of spins~\cite{rummukainen:93,caselle:94,hasenbusch:94,agostini:97}. We shall follow a similar approach, although those authors emphasized equilibrium physics ---our focus, instead, is on non-equilibrium dynamics--- and they were not particularly concerned with massively parallel simulations. 
Lustig's microcanonical ensemble~\cite{lustig:98} also contains a number of daemons proportional to the number of spins.} Lustig's framework is very effective for studying first-order phase transitions~\cite{martin-mayor:07}, also for disordered systems~\cite{fernandez:08}, which suggests a potential effectiveness in the study of spin glasses {(see also~\cite{rummukainen:93} for a study of first-order transitions in non-disordered systems)}. This formalism, however, has three major inconveniences: (i) it is unsuitable for parallel computations, (ii) it is \emph{not} frugal in its usage of random numbers and (iii) it eventually approaches the equilibrium state corresponding to the microcanonical ensemble (which coincides with the canonical ensemble only in the limit of an infinite system size). We cure problems (i) and (ii) by simply skipping the integration of the daemons that Lustig carries out and keeping them instead as explicit dynamic variables. 
In order to cure problem (iii), we generalize the simulated annealing algorithm~\cite{kirkpatrick:83} to the microcanonical context. We argue (and show by example) that it is sufficient to lower the total energy of the system by acting solely on the daemons at a logarithmic time rate (this is the only part of the algorithm that involves random numbers) to obtain results indistinguishable (after a simple time rescaling) from those produced using a canonical simulation employing the Janus II special-purpose computer~\cite{janus:14}. Although rescalability will be considered in much deeper detail below, as a starter, the reader may wish to check Fig.~\ref{fig:E}.

In a canonical simulation, one may define an effective
temperature (for instance) by comparing the instantaneous value of the energy with that of an equilibrated system at the effective temperature. Only in equilibrium does the effective temperature coincide with the thermal-bath temperature. In close analogy, see the more detailed discussion in Sect.~\ref{subsect:ener-annealing}, the refresh of the daemons/walkers effectively cools the system only as long as thermal equilibrium has not been reached. In equilibrium, the net energy flow between spins and daemons/walkers cancels.

The rescalability evinced in Fig.~\ref{fig:E} can be related to analytical work  on microcanonical dynamics ---see, for instance, the Chapter entitled ``Critical Dynamics'' in  Ref.~\cite{zinn-justin:05} for model-C critical dynamics (\emph{i.e.}, non-conserved order parameter and conserved energy). If the coherence length grows with time without bound ---which is the case for a spin glass at $T<T_\mathrm{c}$, $T_\mathrm{c}$ being the critical temperature--- and if the specific heat does not diverge in that limit (which is also the case for spin glasses), one expects that the canonical and the microcanonical dynamics can be superimposed at long times after a simple time rescaling. Although the algorithm proposed here is \emph{not} microcanonical, it alternates between microcanonical and non-microcanonical phases. Hence, the field-theoretical prediction is reassuring. 

The structure of the rest of the paper is as follows: We recall the Edwards-Anderson model and explain our algorithm in Sec.~\ref{sect:model_and_algorithm}. Then, we compare the microcanonical and canonical dynamics for the Edwards-Anderson model in three dimensions in Sec.~\ref{sect:tests}, showing that both belong to the same dynamic universality class. We conclude by summarizing the results and discussing potential future applications of this algorithm in Sec.~\ref{sect:conclusions}. We also provide additional information in three appendices. \ref{ap:time-shift} provides technical details about our time-rescaling procedures, ~\ref{appendix:CUDA} provides some basic information about our publicly available CUDA code. Finally, \ref{appendix:PTgaussian} briefly describes the application of the method in equilibrium simulations for the specific example of a small sample of a three-dimensional spin glass with Gaussian couplings at low temperatures.

\section{The model and the algorithm}
\label{sect:model_and_algorithm}

We have chosen to benchmark our algorithm in the rather demanding context of the three-dimensional Edwards-Anderson model~\cite{edwards:75,edwards:76}, which is the standard for spin glasses. While describing the model in Sec.~\ref{subsect:themodel}, we shall emphasize its separability, which is the crucial feature that makes parallel computation feasible. A similar separation between mutually non-interacting degrees of freedom is a necessary condition in any envisaged application of the solution proposed herein.

We have chosen to organize this section in several separate paragraphs. After the somewhat introductory Secs.~\ref{subsect:themodel} and~\ref{subsect:Lustig}, each paragraph will discuss a specific decision we needed to make to design the algorithm. Indeed, efficient implementations require tuning Monte Carlo algorithms to the user's specific needs, which are not only dictated by the peculiarities of the problem under study, but also by those of the hardware platform that will be used in the simulation. We hope that making our tuning process explicit will help the reader adapt the algorithm to their needs. We have summarized the defining details of the algorithms that we have numerically tested against a canonical simulation in Sec.~\ref{subsect:our-algorithm}.

\subsection{The model (and its separability)}\label{subsect:themodel}

We consider $N$ Ising spins $s_{\boldsymbol{x}}=\pm 1$ that occupy the nodes, denoted by $\boldsymbol{x}$, of a cubic lattice with linear size $L$ ($N=L^3$). The lattice is endowed with periodic boundary conditions. The spins interact with their lattice nearest neighbors through the Hamiltonian
\begin{equation}
\label{eq:Hamiltonian}
H^{\text{\tiny EA}} = - \sum_{\langle \boldsymbol{x},\boldsymbol{y}\rangle} J_{\boldsymbol{x},\boldsymbol{y}} s_{\boldsymbol{x}}s_{\boldsymbol{y}}\,.
\end{equation}
The quenched couplings $J_{\boldsymbol{x},\boldsymbol{y}}$ are independent and identically distributed random variables that are drawn at the beginning of the simulation and kept fixed. The couplings are typically drawn from a symmetric distribution. Our choice will be $J_{\boldsymbol{x},\boldsymbol{y}}=\pm 1$ with 50\% probability (Gaussian-distributed couplings are a fairly popular choice, as well). A specific realization of the coupling matrix $\{J_{\boldsymbol{x},\boldsymbol{y}}\}$ is named a \emph{sample}. With binary couplings $J=\pm 1$, the system undergoes a continuous phase transition at $\Tc = 1.1019(29)$~\cite{janus:13}.

In principle (but see below for an alternative point of view), one is interested in the canonical ensemble, where spin configurations are found as dictated by Boltzmann's probability at inverse temperature $\beta=1/T$:
\begin{equation}\label{eq:canonical}
p(\{s_{\boldsymbol{x}}\})=\frac{\text{e}^{-\beta H^{\text{\tiny EA}}(\{s_{\boldsymbol{x}}\})}}{Z_{\text{\tiny EA}}}\,,\ Z_{\text{\tiny EA}}=\sum_{\{s_{\boldsymbol{x}}\}}\,\text{e}^{-\beta H^{\text{\tiny EA}}(\{s_{\boldsymbol{x}}\})}\,.
\end{equation}
It is fairly common in spin glasses that thermal equilibrium will not be achievable in a reasonable computer time (nor in reasonable human time, if we are talking about a laboratory experiment). Many important experimental effects can still be investigated through an out-of-equilibrium simulation~\cite{janus:17,janus:17b,janus:18,janus:19,janus:21,zhai-janus:20a,janus:23,orbach-janus:23,orbach-janus:24,dahlberg:25}, which is carried out with a local algorithm, such as Metropolis or heat bath (see, for instance, Refs.~\cite{sokal:97,landau:05}) that has Eq.~\eqref{eq:canonical} as its stationary distribution.

A crucial feature of the Hamiltonian $H^\text{\tiny EA}$ that makes parallel computations feasible is separability, which is itself a consequence of the limited interaction range (lattice nearest neighbors in our case). Here, separability means that the $N$ spins can be split {into disjoint sets such that spins in the same set are mutually non-interacting (entailing that they can be updated simultaneously). The number of non-interacting sets could be larger than two, but it should be small to make parallel computing efficient. In our case, a very natural possibility is to group the spins into \emph{even} and \emph{odd} sublattices, depending on the parity of the sum of their Cartesian coordinates. This choice results in a {\em checkerboard} decomposition of the lattice which has proven instrumental in many computational and theoretical studies for decades (Ref.~\cite{barma:77} provides an early example).}

As we discuss in Sec.~\ref{subsect:Lustig}, separability is endangered by the (in so many other ways fully satisfactory) Lustig microcanonical formalism.

\subsection{A generalization of Lustig's microcanonical ensemble}\label{subsect:Lustig}

Initially, we shall follow Lustig and expand the configuration space, introducing new degrees of freedom that we shall name hereafter either daemons or walkers (the distinction is discussed in Sect.~\ref{subsect:wandering}). Specifically, every spin will be endowed with  a number $\gamma$ of these auxiliary degrees of freedom $n_{\boldsymbol{x},\alpha}\geq0$, $\alpha=0,1,2,\ldots,\gamma-1$.
The $n_{\boldsymbol{x},\alpha}$ could be either real or integer.
Our total Hamiltonian will be [$H^\text{\tiny EA}$ was defined in Eq.~\eqref{eq:Hamiltonian}]
\begin{equation}\label{eq:H-total}
    H=H^\text{\tiny EA}\ +\ H^\text{aux}
\end{equation}
with 
\begin{equation}
    H^\text{aux}=\sum_{\boldsymbol{x}}\sum_{\alpha=0}^{\gamma-1}\, n_{\boldsymbol{x},\alpha}\,.
\end{equation}
The corresponding canonical ($p_\text{c}$) and microcanonical ($p_\text{mic}$) probability densities are
\begin{eqnarray}
p_\text{c}(\{s_{\boldsymbol{x}},n_{\boldsymbol{x},\alpha}\};\beta)&=&\frac{\text{exp}[-\beta H(\{s_{\boldsymbol{x}},n_{\boldsymbol{x},\alpha}\}]}{Z_\text{c}}\,,\label{eq:canonical2}\\
p_\text{mic}(\{s_{\boldsymbol{x}},n_{\boldsymbol{x},\alpha}\};{\cal E})&=&\frac{\delta[{\cal E}- H(\{s_{\boldsymbol{x}},n_{\boldsymbol{x},\alpha}\})]}{Z_\text{mic}}\,,\label{eq:Pmic}
\end{eqnarray}
where $\delta$ stands for Dirac's delta function and the partition functions $Z_\text{c}$ and $Z_\text{mic}$ ensure correct normalization as in Eq.~\eqref{eq:canonical}.

If we were to follow Lustig, our next step would be to integrate out the auxiliary degrees of freedom in Eq.~\eqref{eq:Pmic}:
\begin{equation}\label{eq:Pmic2}
p_\text{mic}(\{s_{\boldsymbol{x}}\};{\cal E})=\frac{\e^{(\gamma N -1)\log[{\cal E}- H^\text{\tiny EA}(\{s_{\boldsymbol{x}}\})]}}{\hat Z_\text{mic}}\theta({\cal E}- H^\text{\tiny EA}[\{s_{\boldsymbol{x}}\})]\,,
\end{equation}
where the Heaviside step function ensures that ${\cal E}>H^\text{\tiny EA}(\{s_{\boldsymbol{x}})\}$, and the modified partition function $\hat Z_\text{mic}$ ensures normalization.  Eq.~\eqref{eq:Pmic2} is well adapted for a serial Monte Carlo simulation, where spins are updated one by one (an independent random number would be necessary per every spin update). Unfortunately, the effective action in Eq.~\eqref{eq:Pmic2}, $\log({\cal E}-H_{\text{\tiny EA}})$ is highly non-linear in $H^\text{\tiny EA}$, which spoils the  separability property discussed in Sec.~\ref{subsect:themodel}. Indeed, even an effective action quadratic in $H^\text{\tiny EA}$ would be non-separable because \emph{every} pair of spins would be mutually interacting. With some ingenuity, a limited parallel computation can be set for an all-to-all interaction~\cite{aramon:19}, but we regard this as a very undesirable feature. Hence,  we shall depart from Lustig's approach at this point.

\subsection{Creutz's move saves the day}\label{subsec:creutzmove}

Fortunately, the obstacle described in the previous paragraph can be cleared by adapting to our needs the basic move from Creutz's algorithm~\cite{creutz:83}. In this way, three different problems get solved at once: (i) we obtain an algorithm that verifies detailed balance for both the canonical and the microcanonical probabilities in Eqs.~\eqref{eq:canonical2} and~\eqref{eq:Pmic}, (ii) the move does not require any random number, and (iii) separability is not endangered (hence parallel simulations are feasible).

Let us describe the elementary move of our modified form of Creutz's algorithm. Pick spin $s_{\boldsymbol{x}}$ and one of its daemons/walkers, $n_{\boldsymbol{x},\alpha}$, and compute the change in $H^\text{\tiny EA}$, Eq.~\eqref{eq:Hamiltonian}, if we flip the spin at site $\boldsymbol{x}$ (that is, $s_{\boldsymbol{x}}\to -s_{\boldsymbol{x}}$) while keeping all other spins fixed, $\Delta H^\text{\tiny EA}$. If we update
\begin{equation}\label{eq:Creutz}
n_{\boldsymbol{x},\alpha} \longrightarrow n_{\boldsymbol{x},\alpha}-\Delta H^\text{\tiny EA}\,,
\end{equation}
the total Hamiltonian~\eqref{eq:H-total} remains unchanged because the total energy gained (or lost) by the spin system is exactly compensated by the $n_{\boldsymbol{x},\alpha}$. However, to obtain a valid new configuration, we should have $n_{\boldsymbol{x},\alpha}\geq 0$ after the update in Eq.~\eqref{eq:Creutz}. The spin is flipped, and the $n_{\boldsymbol{x},\alpha}$ is updated only if the new configuration is legal. Up to now, the only difference from Creutz lies in that he worked with a single auxiliary degree of freedom that is shared by the whole system. In our Lustig-inspired setting, every spin has its own set of $\gamma$ daemons/walkers $n_{\boldsymbol{x},\alpha}$.

Let us now briefly explain why our elementary move solves the three problems mentioned above.

The proof of detailed balance ---for both the canonical~\eqref{eq:canonical2} and the microcanonical weight~\eqref{eq:Pmic}--- lies in the reversibility of our elementary move. Indeed, if we pick exactly the same spin $s_{\boldsymbol{x}}$ and $n_{\boldsymbol{x},\alpha}$
\emph{after} we carried out the elementary move ---all other spins and daemons/walkers unchanged--- and try to update them again, we would exactly undo the spin flip and the update~\eqref{eq:Creutz} because $\Delta H^\text{\tiny EA}$ changes sign after the spin flip. And, of course, the newly reached configuration (which is the configuration we started with) would be legal.
Using the language of the Metropolis algorithm~\cite{sokal:97}, we have a symmetric proposal of change in the configuration space that leaves the total energy invariant [implying that the probabilities for the starting and the final configurations are identical, either in the canonical, Eq.~\eqref{eq:canonical2}, or in the microcanonical ensemble, Eq.~\eqref{eq:Pmic}]. If such a move is accepted with probability one, detailed balance is fulfilled.

Of course, no random numbers are involved in the elementary move. However, it is amusing to notice that if a fresh value of the $n_{\boldsymbol{x},\alpha}$ were drawn for every spin-flip attempt,\footnote{This is easy to do,  although it would require a random number per spin update, because in the canonical ensemble, Eq.~\eqref{eq:canonical2}, the $n_{\boldsymbol{x},\alpha}$ are independent exponentially-distributed random variables.} our elementary move would become the
textbook Metropolis update ---see, e.g., Refs.~\cite{sokal:97,landau:05}--- for the spin system, recall Eq.~\eqref{eq:canonical}. Baroque as it may seem, this disguising of the textbook Metropolis algorithm has resulted in a very efficient multispin coding implementation~\cite{ito:90}.

As for separability, notice that $\Delta H^\text{\tiny EA}$ depends only on spins with parity opposite to the parity of our site $\boldsymbol{x}$. Hence, if we consider two (say) even sites $\boldsymbol{y}$ and $\boldsymbol{z}$, the two determinations on whether $s_{\boldsymbol{y}}$ and $s_{\boldsymbol{z}}$ can be flipped depend \emph{solely} on odd spins and their chosen $n_{\boldsymbol{y},\alpha_y}$ and $n_{\boldsymbol{z},\alpha_z}$. If the spins at odd sites are held fixed, it is immaterial which of the two sites, $\boldsymbol{y}$ or $\boldsymbol{z}$, is updated first. A parallel update of all sites with a common lattice parity is, then, feasible.

\subsection{Continuous or discrete daemons/walkers?}

A computationally convenient feature of the $J=\pm 1$ choice in $H^\text{\tiny EA}$, Eq.~\eqref{eq:Hamiltonian}, is that the energy changes that a single spin-flip may cause are limited. In three spatial dimensions, $\Delta H^\text{\tiny EA}=0,\pm 4, \pm 8$ or $\pm 12$.  Hence, in our particular case, it is practical to limit the possible values of the daemons/walkers to integers multiple of four $n_{\boldsymbol{x},\alpha}=4m\,,\ m=0,1,2,3,\ldots$. For Gaussian couplings, it would probably be more natural to keep continuous daemons/walkers, or have them take values on a finer mesh of energies.

 In the canonical ensemble \emph{only}, the $\gamma N$ auxiliary variables $n_{\boldsymbol{x},\alpha}$ are statistically independent from each other and from the spins. In the continuous case, the $n_{\boldsymbol{x},\alpha}$ are exponentially distributed:
\begin{equation}
\text{Prob}(0 < n_{\boldsymbol{x},\alpha}  < R)= 1 - \exp(-\beta R).
\end{equation}  
In Lustig's original ensemble the  $n_{\boldsymbol{x},\alpha}$ enter quadratically in the $H^\text{aux}$~\cite{lustig:98}, so that their distribution is Gaussian. For the case with discrete daemons/walkers, see Eq.~\eqref{eq:prob-daemons-walkers} below.

\subsection{Daemons/walkers with a limited capacity}
\label{subsec:limitedcapacity}

To further parallelize our computation we employ multispin
coding~\cite{jacobs:81}. We use the different bits in a memory word to
represent different physical spins. Hence, we have chosen to restrict
the possible values to $n_{\boldsymbol{x},\alpha}=4m$ with
$m=0,1,2,3$. Thus $m$ is representable with two bits only, and we can
use two 32-bit words to represent 32 different
daemons/walkers. Note that these choices are
  problem specific. In other situations, different choices are possible, see, for instance, \ref{appendix:PTgaussian}.

However, this limitation in the maximum value of $n_{\boldsymbol{x},\alpha}$ comes with costs and benefits. On the one hand, the configuration obtained after Creutz's move will be legal, meaning that the spin flip will be accepted, if and only if
\begin{equation}\label{eq:flip-OK}
0\leq n_{\boldsymbol{x},\alpha}-\Delta H^\text{\tiny EA}\leq 12\,.
\end{equation}
In other words, we should \emph{reject} any spin-flip that lowers the energy by an amount larger than what the daemon/walker responsible for that update can currently accommodate. On the other hand, having daemons/walkers representable with just two bits reduces the total memory needed. Furthermore, in a multispin coding simulation, the whole algorithm should be implemented employing solely Boolean operations [this includes checking condition  Eq.~\eqref{eq:flip-OK} or performing the update in Eq.~\eqref{eq:Creutz}], which is certainly simpler if the daemons/walkers are representable with only two bits. {Refs.~\cite{caselle:94,hasenbusch:94,agostini:97} also considered daemons with limited capacity and managed them accordingly.}

Daemons/walkers will be initialized ---or refreshed--- according to their canonical weight, Eq.~\eqref{eq:canonical2}. That is to say, as independent, identically distributed random variables drawn from the distribution
\begin{equation}\label{eq:prob-daemons-walkers}
p(n=4m)=\e^{-4m\beta}\frac{1-\e^{-4\beta}}{1-\e^{-16\beta}}\,,\ 
m=0,1,2,3\,.
\end{equation}
We conclude this paragraph with a word of caution. The daemons/walkers should be allowed to take values at least as large as the largest possible $|\Delta H^\text{\tiny EA}|$ (the problem is in the negative values of $H^\text{\tiny EA}$). In this way, it will be ensured that no spin (or the equivalent elementary variable) will lock forever at local energy minima too deep to be evaded from. 

\subsection{Daemons or walkers?}\label{subsect:wandering}
One might worry about the presence of local conservation laws. Local energy conservation is needed if we want to use a parallel updating algorithm, but it could cause dynamic slowing down. This is why we have considered two variants of the algorithm. 

If the auxiliary variable $n_{\boldsymbol{x},\alpha}$ is forever bound to a lattice site $\boldsymbol{x}$, it will be named a \emph{daemon}. Yet, if it is allowed to wander through the lattice, it will be called a \emph{walker}.

We have designed a wandering procedure that runs efficiently on a GPU because it can be implemented using a memory-access pattern that fulfills the memory coalescing conditions. Consider one of the lattice directions (say Y) and a moving direction (backward or forward). A forward-Y wandering is a two-timestep cycle that can be implemented only for even lattice size $L$:
\begin{enumerate}
    \item At the starting time of the cycle, all the $\alpha$-th walkers sitting on even lattice sites exchange their values with the $\alpha$-th walkers sitting on their lattice nearest neighbors along the positive Y direction (periodic boundary conditions are employed; the neighbor will be an odd site).
     \item At the second timestep in the cycle, the walkers at odd lattice sites will exchange their values with their lattice nearest-neighbor along the positive Y direction.
\end{enumerate} 
Note that after $L/2$ cycles, or $L$ timesteps, every forward-Y wandering walker will return to its starting point, although it will have been modified along its way through the updates in Eq.~\eqref{eq:Creutz}.

The backward-Y wandering operation differs from the forward-Y counterpart just in that the lattice nearest-neighbor will be sought along the \emph{negative} Y direction.

{Interestingly enough, Refs.~\cite{caselle:94,hasenbusch:94} implemented their own daemon wandering (although they refreshed their daemons far more often than we do, see Sect.~\ref{subsect:ener-annealing}).}
\subsection{The microcanonical simulated annealing}\label{subsect:ener-annealing}
The operations described so far do not allow us to approach the equilibrium canonical distribution, Eq.~\eqref{eq:canonical2}. Indeed,  with our microcanonical update, the total energy ${\cal E}$ remains equal to its starting value ${\cal E}_\text{start}$. Eq.~\eqref{eq:H-total} tells us that ${\cal E}_\text{start}$ is the sum of two contributions, the starting value of the spin Hamiltonian $H^\text{\tiny EA}$~\eqref{eq:Hamiltonian} ---typically much larger than the equilibrium value at inverse temperature $\beta, \langle H^\text{\tiny EA}\rangle_\beta$--- and the energy of the daemons/walkers. Although the daemons/walkers will be directly initialized in thermal equilibrium through Eq.~\eqref{eq:prob-daemons-walkers}, it is clear that ${\cal E}_\text{start}$ will be too high. As a result, the microcanonical updates will lower the energy of the spins---which is desirable--- at the cost of heating the daemons/walkers ---which is undesirable. It is easy to assign an instantaneous temperature to the daemons/walkers, because the $\beta$-dependent equilibrium expectation value  can be easily computed from Eq.~\eqref{eq:prob-daemons-walkers}. We have $\gamma$ daemons/walkers per each of the $N$ spins, as
\begin{equation}\label{eq:daemons-temperature}
\frac{\langle H^\text{aux}\rangle}{\gamma N}=\frac{4\,\e^{-4\beta}}{1-\e^{-4\beta}}-\frac{16\,\e^{-16\beta}}{1-\e^{-16\beta}}\;.
\end{equation}
Hence, one may simply invert Eq.~\eqref{eq:daemons-temperature} to obtain the effective temperature from the instantaneous value \changes{of the daemon/walker energy}. Since the microcanonical dynamics will increase the value of $H^\text{aux}$ by relaxing the spin system, one may aptly describe the process by saying that the effective temperature of the daemons/walkers is raised because of the energy absorbed from the spins.

Now, if we aim to reach the canonical equilibrium, Eq.~\eqref{eq:canonical2}, it is clear that we need to find a way to \emph{cool} the daemons/walkers. In some ways, our solution will be similar to the strategy known as simulated annealing~\cite{kirkpatrick:83}. At specified times, see the discussion below, we throw away the daemon/walker configuration ---leaving spins untouched--- and generate a new, statistically independent one from Eq.~\eqref{eq:prob-daemons-walkers}. By construction, the \changes{new daemon/walker configuration} will have the correct effective temperature dictated by Eq.~\eqref{eq:daemons-temperature}. In other words, the \changes{nominal daemon/walker temperature $T$} will coincide with the effective temperature. Hence, we shall have decreased the total energy ${\cal E}$ of our \emph{annealing} system. When equilibrium is finally reached, the energy gains and losses suffered by the spins finally compensate, and, although the effective temperature of the daemons/walkers fluctuates, it remains stable (on average).

From the above discussion, it should be clear that refreshing \changes{the daemon/walker configuration} at regular times, say every 1000 timesteps (see Sec.~\ref{subsect:our-algorithm} for the precise definition of timestep), should eventually lead any system of finite size to canonical equilibrium. This would dilute the need for random numbers by a factor of 1000, which would probably make the computational weight of generating random numbers irrelevant for most (if not all) applications. However, we shall try to be even thriftier with our random numbers. Indeed, the energy relaxation of spin glasses at and below their critical temperature decreases very slowly with time $t$. The excess energy over the equilibrium value decays as a power law, 
$H^\text{\tiny EA}(t) -\langle H^{\text{\tiny EA}}\rangle_\beta\propto N/t^{\epsilon(T)}$, 
where $\epsilon(T)$ is a fairly small exponent~\cite{janus:09b,janus:19} (the exponent is so small that the power law resembles a logarithmic behavior). In fact, for \changes{a $1/\log (t)$ law}, the time required to have a prefixed energy factor decay $f$ grows exponentially with $1/f$.  We decided accordingly to refresh our daemons/walkers \changes{with a schedule that is linear with the logarithm of the time}.  
At each of those timesteps, the spin system will experience an abrupt, but tiny, energy change as a reaction to the cooler daemons/walkers. This is inappreciable at the scale of Fig.~\ref{fig:E}, but can be observed with the more refined analysis in Fig.~\ref{fig:T0.7}-d.

\subsection{Putting all pieces together: the algorithms we tested}\label{subsect:our-algorithm}

We have tested the algorithm described above with $\gamma=1,2,4$ and 6 daemons per spin. We have also tried the one with $\gamma=1$ and $6$ walkers per spin.

In all cases, a timestep consisted of $\gamma$ full-lattice sweeps (\emph{i.e.}, the flip of every spin was considered $\gamma$ times within a single timestep). A sweep consists of updating the $N/2$ sites with even lattice parity, followed by the update of the $N/2$ sites with odd parity. On the $\alpha$-th sweep in a timestep ($\alpha=0,1,\ldots\gamma-1$), the daemon/walker $n_{\boldsymbol{x},\alpha}$ was used in the update of the site $\boldsymbol{x}$.

When applied, the walker-wandering part of the algorithm occurred once every timestep (the parity of the time index $t$ defines which of the two wandering operations is used, recall Sec.~\ref{subsect:wandering}). In the case with just  $\gamma=1$ walker per spin, we employed forward-X wandering. In the case with $\gamma=6$ walkers per spin, we used
\begin{equation}
\begin{split}
\alpha=0\quad &\longrightarrow\quad  \text{forward-X}\,,\\
\alpha=1\quad &\longrightarrow\quad  \text{forward-Y}\,,\\
\alpha=2\quad &\longrightarrow\quad  \text{forward-Z}\,,\\
\alpha=3\quad &\longrightarrow\quad  \text{backward-X}\,,\\
\alpha=4\quad &\longrightarrow\quad  \text{backward-Y}\,,\\
\alpha=5\quad &\longrightarrow\quad  \text{backward-Z}\,.
\end{split}
\end{equation}

In all cases, the refresh of the daemons/walkers took place at timesteps $t_s=2^s$, as specified in Sec.~\ref{subsect:ener-annealing}. Observe that our definition of timestep ensures that the number of sweeps that a particular subfamily of daemons/walkers, $\alpha$, completes between consecutive refreshing steps is independent of $\gamma$.

\subsubsection{The algorithm in the form of a procedure}

\begin{enumerate}
    \item Initialize $s=0$ and the time counter variable $t=0$.
    \item Get a fresh daemons/walkers configuration $\{n_{\boldsymbol{x},\alpha}\}$ from  Eq.~\eqref{eq:prob-daemons-walkers}.
    \item  Carry out $2^s$ \emph{steps}. Each step consists of:
    \begin{enumerate}
        \item Cycle for $\alpha=0,\ldots,\gamma-1$:
        \begin{enumerate}
            \item For every site in the even sublattice, $\boldsymbol{x}$, compute $\Delta H^\text{\tiny EA}$ for the spin-flip of $s_{\boldsymbol{x}}$ and use $n_{\boldsymbol{x},\alpha}$ to check condition in Eq.~\eqref{eq:flip-OK}. If the condition is met, flip 
            the spin and update the daemon/walker as in Eq.~\eqref{eq:Creutz}.
            \item Same as (i) for the odd sublattice.
        \end{enumerate}
        \item If the $\{n_{\boldsymbol{x},\alpha}\}$ are walkers, carry out the part of the wandering cycle dictated by the parity of $t$, see Sect.~\ref{subsect:wandering} (for daemons, do nothing).
        \item Increase $t$ by one unit.
    \end{enumerate}
    \item Increase $s$ by one unit and go back to (2).
\end{enumerate}
We insist on updating all sites with a given parity first, and then the sites with the opposite lattice parity, just because we intend to carry out a parallel computation. Readers willing to work with a serial update may access the lattice in whatever order they prefer.

\section{Results}
\label{sect:tests}

\begin{figure*}[t]
    \centering
    \includegraphics[width=.95\linewidth]{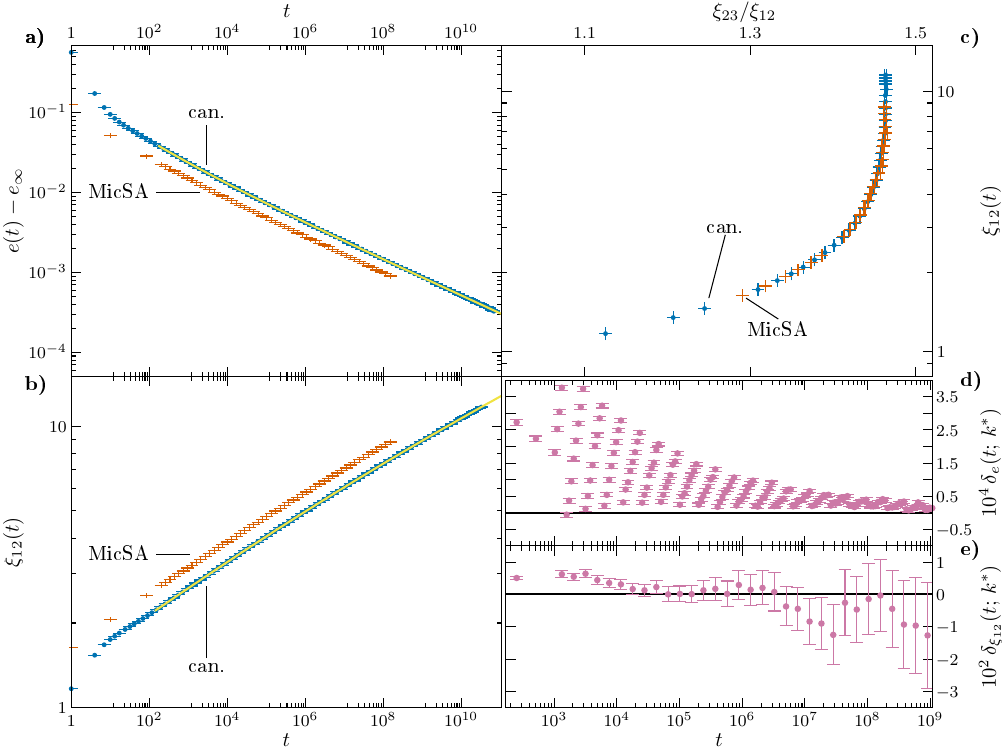}
    \caption{{Comparison of Metropolis and MicSA dynamics with 6 walkers in the spin-glass phase at $T\!=\!0.7\approx 0.64 T_\mathrm{c}$}. \textbf{a)} Time evolution of the difference of the energy at time $t$ and the equilibrium value obtained through a fit to the canonical data. The solid yellow line is for the best fit described in Table~\ref{tab:fit-info}. \textbf{b)} The coherence length, $\xi_{12}$, as a function of the simulation time for both models. The solid line is again for the best fit described in Table~\ref{tab:fit-info}. \textbf{c)} Coherence length, \changes{$\xi_{12}(t)$}, as a function of the ratio between the coherence lengths \changes{$\xi_{23}(t)$} and \changes{$\xi_{12}(t)$}. \textbf{d)} Difference between the results from the two algorithms  for the energy, Eq.~\eqref{eq:delta1},  as a function of the simulation time. \textbf{e)} As in \textbf{d)} but for the coherence length. To compute these differences, we have considered the best fit to our canonical data ---solid lines in \textbf{a)} and \textbf{b)}--- and the rescaling coefficient $k^*=6.034$ (see \ref{ap:time-shift} for more information on the estimation of $k^*$). Error bars are smaller than data points. For the sake of clarity points are plotted at regular time intervals in logarithmic scale. \label{fig:T0.7}}
\end{figure*}

\begin{figure*}[t]
    \centering
    \includegraphics[width=.95\linewidth]{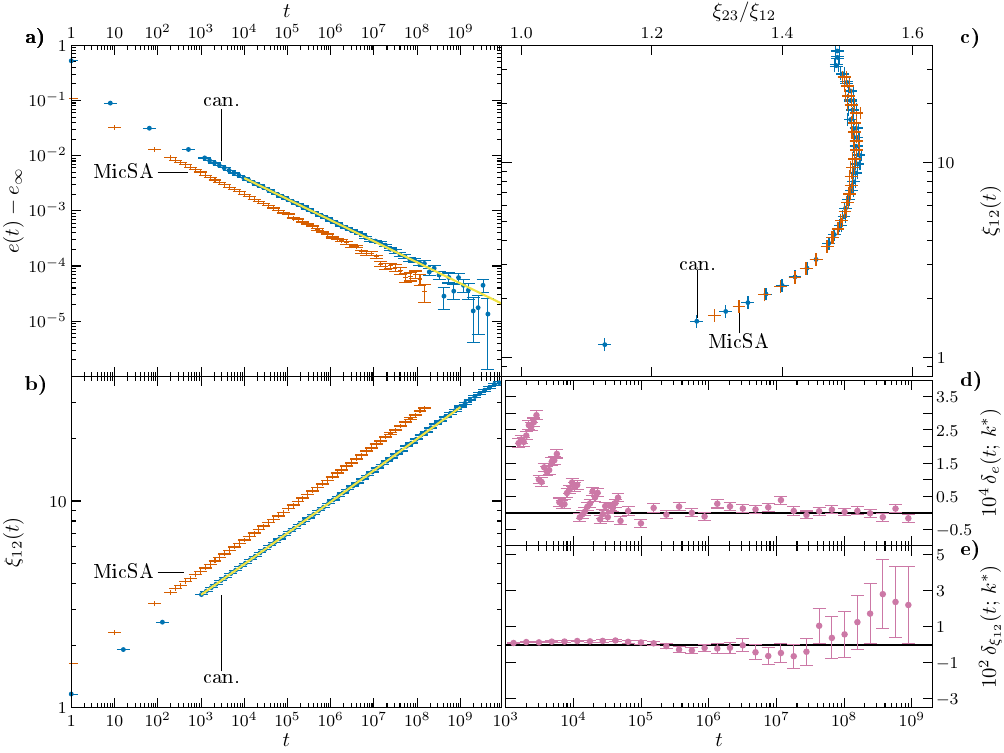}
    \caption{{Six walkers at  $T=1.1\simeq T_\mathrm{c}$}. As in Fig.~\ref{fig:T0.7}, but with the time-rescaling factor $k^*=5.876$ (see Table~\ref{tab:k*}).\label{fig:T1.1}}
\end{figure*}

\begin{figure*}[t]
    \centering
    \includegraphics[width=.95\linewidth]{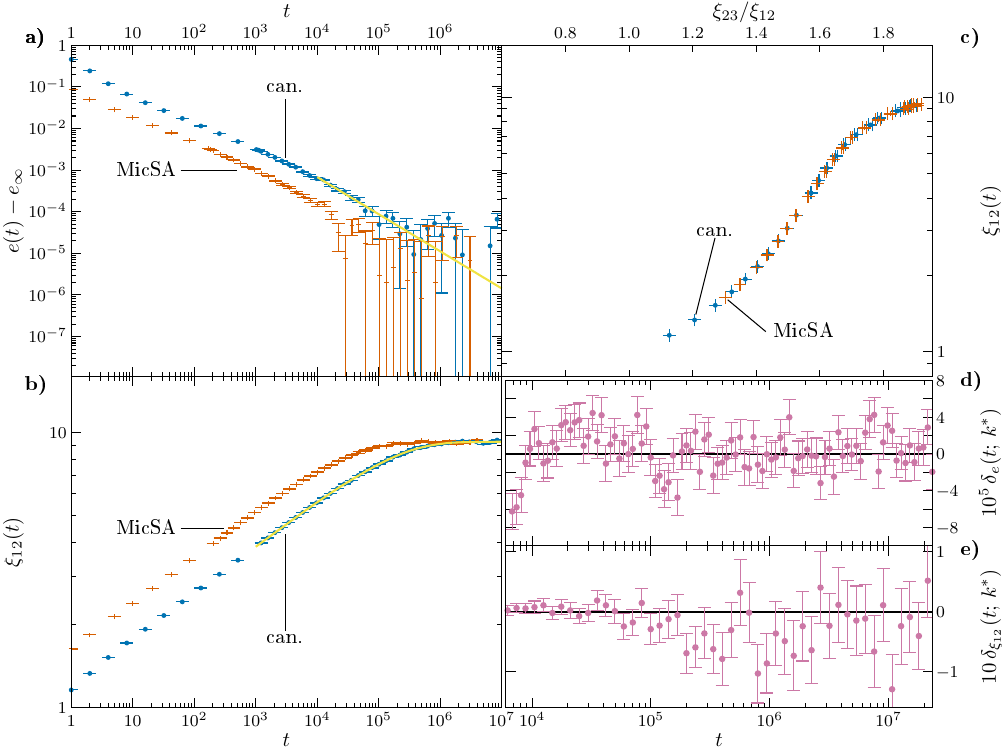}
    \caption{{Six walkers in the paramagnetic phase at $T\!=\!1.4\approx1.27\Tc$}. As in Fig.~\ref{fig:T0.7}, but with the time-rescaling factor  $k^*=5.904$ (see Table~\ref{tab:k*}). \label{fig:T1.4}}
\end{figure*}

\renewcommand{\arraystretch}{1.5}
\begin{table}[t]
    \centering
    \footnotesize
    \caption{{Estimation of the best time-rescaling factor $k^*$ } as obtained from a fit to Eq.~\eqref{eq:k-def} for the different number of daemons/walkers per spin $\gamma$ and temperatures (we mark with a symbol $^\dagger$ the algorithm based on walkers and not on daemons). \changes{We denote the statistical uncertainty by $\Delta k^*$}. We also report the fits figure of merit $\chi^2/\mathrm{d.o.f}$ as computed from the diagonal part of the covariance matrix (d.o.f. meaning degrees of freedom). The choice of using either daemons or walkers (and how many) results only in a rescaling of the time units, but not in a change of the dynamical critical exponents.
    \label{tab:k*}}
    \begin{tabular}{c c c c c c c }\hline
        $T$ & $\gamma$ & $k^*$ & $\Delta k^*$ & $[\xi_\mathrm{min},\xi_\mathrm{max}]$ & $\chi^2/\mathrm{d.o.f.}$  \\\hline
        $0.7$ & $6^\dagger$ & 6.034 & 0.007 & [4, 12.5] & 34.1/108  \\
        $\Tc$ & 1 & 0.3984 & 0.0012 & [8.5, 14] & 25.8/33 \\
        $\Tc$ & $1^\dagger$ & 0.967 & 0.003 & [8, 12] & 15.4/25 \\
        $\Tc$ & 2 & 1.051 & 0.002 & [7, 12] & 21.3/36 \\
        $\Tc$ & 4 & 2.667 & 0.004 & [6, 14] & 13.2/59 \\
        $\Tc$ & 6 & 4.79 & 0.04 & [18, 30] & 8.4/31  \\
        $\Tc$ & $6^\dagger$ & 5.876 & 0.011 & [7, 30] & 66.0/106  \\
        $1.4$ & $6^\dagger$ & 5.904 & 0.014 & $t>10^3$ & 84.5/95 \\ \hline
    \end{tabular}
\end{table}

We have tested the new set of algorithms under three very different dynamical conditions. In all three cases, we start the simulation from a fully disordered configuration (which mimics infinite temperature). At the starting time, $t=0$, the system is instantaneously quenched to the working temperature $T$, where it is allowed to relax. The features of the relaxation are very much dependent on $T$. In the paramagnetic phase, $T>T_\mathrm{c}$, thermal equilibrium can eventually be reached. Instead, equilibrium is unreachable in a large lattice if one works at the critical point $T=T_\mathrm{c}$, or in the spin-glass phase $T<T_\mathrm{c}$. This impossibility to equilibrate is, exactly, the situation encountered in experiments~\cite{dahlberg:25}. For $T\leq T_\mathrm{c}$ the dynamics consists in the endless growth of glassy magnetic domains of linear size $\xi(t,T)$ (when comparing data at the same temperature we shall use the lighter notation $\xi(t)$; a similar convention will be used for other quantities). The time growth of the coherence length $\xi(t,T)$ turns out to be excruciatingly slow, and slower the lower $T$ is. For $T>T_\mathrm{c}$, instead, $\xi(t,T)$ eventually saturates to its equilibrium value $\xi_\text{eq}(T)$. In order to be representative of the experimental situation, one should have $L\gg\xi(t,T)$ at all times~\cite{janus:08b}. We shall not feel constrained by the  $L\gg\xi(t,T)$ condition, because we want to test our algorithms also when $L$ and $\xi(t,T)$ become comparable.

To investigate the domain-growth process briefly described above, it is mandatory to use replicas, namely statistically independent system copies $\{s_{\boldsymbol{x}}^{(a)}\}$ that evolve under the very same Hamiltonian $H^\text{\tiny EA}$, Eq.~\eqref{eq:Hamiltonian}, \emph{i.e.}, with the same coupling matrix $\{J_{\boldsymbol{x},\boldsymbol{y}}\}$. The superscript $a=1,2,\ldots,N_R$ labels the $N_R$ replicas in a simulation.  For any quantity, $O$, we shall use $\langle O\rangle$ for its average over the initial conditions. In particular, for different replica indices $a\neq b$, one has 
\begin{equation}
    \langle s_{\boldsymbol{x}}^{(a)}(t) s_{\boldsymbol{y}}^{(a)}(t) s_{\boldsymbol{x}}^{(b)}(t) s_{\boldsymbol{y}}^{(b)}(t) \rangle= \langle s_{\boldsymbol{x}}(t) s_{\boldsymbol{y}}(t) \rangle^2\,,
\end{equation} 
because the different replicas are independent, identically distributed random variables.

The main quantities we have used to investigate the dynamics are the energy per spin $e(t)$ and the correlation function $C_4(\boldsymbol{r},t)$. For both quantities, we give the formal definition and, after the $\approx$ sign, the estimators that one can compute with a finite number of replicas $N_R$:
\begin{eqnarray}
e(t)&=& \frac{1}{N}\langle H^\text{\tiny EA}(\{s_{\boldsymbol{x}}(t)\})\rangle\,,\\
&\approx& \frac{1}{N N_R}\sum_{a=1}^{N_R} H^\text{\tiny EA}(\{s_{\boldsymbol{x}}^{(a)}(t)\})\,,\nonumber\\
C_4(\boldsymbol{r},t)&=&\frac{1}{N}\sum_{\boldsymbol{y}-\boldsymbol{x}=\boldsymbol{r}}
\langle s_{\boldsymbol{x}}(t) s_{\boldsymbol{y}}(t) \rangle^2\,,\\
&\approx&\frac{1}{\tilde N}\sum_{a\neq b} \sum_{\boldsymbol{y}-\boldsymbol{x}=\boldsymbol{r}}
s_{\boldsymbol{x}}^{(a)}(t) s_{\boldsymbol{y}}^{(a)}(t) s_{\boldsymbol{x}}^{(b)}(t) s_{\boldsymbol{y}}^{(b)}(t)\,,\nonumber
\end{eqnarray}
where the normalizing factor $\tilde N=N N_R(N_R-1)/2$ takes into account the different ways of choosing two different replica indices. We also have employed the periodic boundary conditions when computing $\boldsymbol{y}-\boldsymbol{x}$. Rotational invariance~\cite{janus:09b} suggests choosing the displacement vector $\boldsymbol{r}$ along one of the lattice axes. In fact, we average over the three different choices $\boldsymbol{r}=(r,0,0), (0,r,0)$ or $(0,0,r)$ and name the result simply $C_4(r,t)$.

Both $e(t)$ and $C_4(r,t)$ are self-averaging quantities, which means that their sample-to-sample fluctuations vanish for large $L$ as a power of the ratio $\xi(t)/L$ ---recall that a sample is a realization of the coupling matrix $\{J_{\boldsymbol{x},\boldsymbol{y}}\}$ in Eq.~\eqref{eq:Hamiltonian}. In fact, experiments are typically performed on a single sample. We have decided to mimic this approach and work with just one sample, obviously the same one for all algorithms (a sample already studied in previous investigations~\cite{janus:18,janus:19,zhai-janus:20a,janus:23,orbach-janus:23,orbach-janus:24,janus:24}), because our target is obtaining very precise comparisons. Eliminating the noise induced by the sample-to-sample fluctuations helps us achieve this purpose and provides a much stronger test of the equivalence between formalisms.

However, while comparing the curves $e(t)$ from different algorithms is relatively straightforward, there is too much information in the $C_4(r,t)$. To simplify the analysis, we squeeze the $r$-dependence of the correlation function into two numbers, the coherence length and a dimensionless ratio of characteristic lengths. Specifically, we follow Refs.~\cite{janus:08b, janus:09b} and compute the integrals
\begin{equation}
I_n=\int_0^\infty\mathrm{d}r\, r^nC_4(r,t)\,.
\end{equation}
The integral converges because at large distances we have $C_4(r,t)\sim f(r/\xi(t))/r^\theta$, where $f(x)$ is a cut-off function decaying super-exponentially in $x$. For technical details about estimating the integrals $I_n$, we refer the reader to Refs.~\cite{fernandez:19,janus:18}. The characteristic lengths $\xi_{n,n+1}$ are computed from the $I_n$,
\begin{equation}\label{eq:xi-def}
    \xi_{n,n+1}=\frac{I_{n+1}}{I_n}
\end{equation}
We shall follow the (by now) standard practice of identifying the spin-glass coherence length with $\xi_{12}(t)$. As for the overall $r$ dependence of $C_4(r,t)$ we consider the dimensionless ratio $\xi_{23}/\xi_{12}$~\cite{fernandez:15,fernandez:18b}. $C_4(r,t)$ is expected to reach a scale-invariant limit at long times [\emph{i.e.}, large $\xi_{12}(t)$] when $T\leq T_\mathrm{c}$, which implies that the ratio $\xi_{23}/\xi_{12}$ will reach a long-time limit as well~\cite{janus:18}.

Our estimates for $e(t)$, $\xi_{12}(t)$ and the dimensionless ratio $\xi_{23}/\xi_{12}(t)$ are displayed as a function of time in Fig.~\ref{fig:T0.7} ($T=0.7\approx 0.64 T_\mathrm{c}$), Fig.~\ref{fig:T1.1} ($T=1.1\approx T_\mathrm{c}$) and Fig.~\ref{fig:T1.4} ($T=1.4\approx 1.27 T_\mathrm{c}$). For all three temperatures, we observe that, when $\xi_{12}(t)$ is plotted parametrically as a function of $\xi_{23}/\xi_{12}(t)$, the canonical and the MicSA data fall onto a single curve. This finding indicates that it suffices to consider $\xi_{12}(t)$ when trying to find the correspondence between the $C_4(r,t)$ computed with the two dynamics. The figures suggest as well that a shift in $\log t$  (\emph{i.e.}, a simple rescaling for $t$) suffices to superimpose the MicSA data with the canonical curve for both  $e(t)$ and $\xi_{12}(t)$. We shall attempt to make this observation quantitatively precise.

The technical difficulty to address is that $e(t)$ and $\xi_{12}(t)$ are known only on a discrete time grid. We overcome this difficulty by interpolating the curves from the canonical simulations with a simple, physically motivated continuous function of $t$. The interpolating function is found from a fit to
\begin{equation}\label{eq:F-interpol}
    \xi^\text{can}_{12}(t,T)=F^\xi_\text{can}(t,T)\,,\quad e^\text{can}(t,T)=F^\e_\text{can}(t,T)\,.
\end{equation}
The form of the interpolating functions varies very much with temperature, as explained in \ref{ap:time-shift}, where details on the above fits are provided (in particular, for $T\leq T_\mathrm{c}$ it is simpler to find $t$ as a function of $\xi_{12}$ than the other way around). The reader should take neither $F^\xi_\text{can}(T,t)$ nor $F^e_\text{can}(T,t)$ too far, because they are mere interpolating devices that are not good for extrapolations. Among other problems, these functions represent the results of a single sample, including times when ---for $T\geq T_\mathrm{c}$--- the condition $L\gg\xi(T,t)$ is no longer met.

\begin{figure}[tb]
    \centering
    \includegraphics[width=\linewidth]{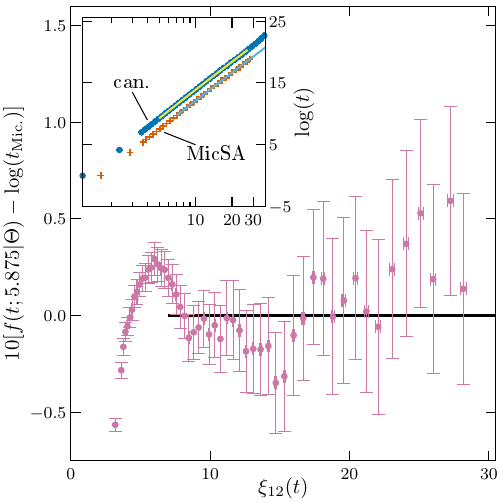}
    \caption{{Estimation of the best time shift $k^*$
        at \Tc}. In the main panel, the difference between the
      logarithm of the original measured time in the MicSA data and
      the interpolating function $f(k^*t)$ as a function of the
      coherence length $\xi_{12}$. The horizontal solid line at zero
      indicates the interval of the fit used to estimate $k^*$ (see
      Table~\ref{tab:k*} for details). {Inset:} $\log(t)$ as a
      function of $\xi_{12}$ for both algorithms. Solid lines indicate
      the best fit for our data. In the case of the MicSA data, the
      only fitting parameter is $k$, while the rest of the parameters
      $\Theta=\lbrace a_0,\,a_1,\, z\rbrace$ come from the fit to the
      canonical data described in \ref{ap:time-shift}.}
    \label{fig:fit_k*}
\end{figure}

Having promoted \changes{$\xi_{12}^{\text{can}}(t,T)$} to a continuous function, we can simply fit the MicSA curves to
\begin{equation}\label{eq:k-def}
     \xi^\text{MicSA}_{12}(t,T)=F^\xi_\text{can}(k\,t,T)\,,
\end{equation}
with the rescaling factor $k$ as the only fitting parameter. The best-fit rescaling parameter is named $k^*$ in Table~\ref{tab:k*} and in Figs.~\ref{fig:T0.7},~\ref{fig:T1.1}, and~\ref{fig:T1.4}.  In the figures, we have used for the energy the same value of $k^*$ computed with $\xi_{12}$. Each  of the different variants of the MicSA algorithm will have its own $k^*$ that slightly depends on the temperature. For further details on the computation of $k^*$, see Fig.~\ref{fig:fit_k*} and \ref{ap:time-shift}.

\begin{table*}[t]
\caption{Results of the different fits to our canonical data shown in
  Figs.~\ref{fig:E}-~\ref{fig:T1.4}. These fits correspond to the
  interpolation used to estimate $\delta_e(t,k)$ and $\delta_\xi(t,k)$
  [see Eqs.~\eqref{eq:delta1} and~\eqref{eq:delta2}] explained in
  Sec.~\ref{sect:tests}. \changes{We specify the time interval considered in the fits as $[t_{\mathrm{min}}, t_{\mathrm{max}}]$ ($*$ indicates that all times $t\geq t_{\mathrm{min}}$ are  included in the fit).}
    \label{tab:fit-info}}
{\footnotesize
    \begin{tabular}{ c c c c c c }
    \hline
        $\mathcal{O}$ & $T$ & \changes{$F^{\#}_{\mathrm{can}}(t)$} & $[t_{\mathrm{min}}, t_{\mathrm{max}}]$ & Fit & $\chi^2/\mathrm{d.o.f.}$ \\\hline 
    $e$ & $0.7$ & $e_\infty+a_1t^{-z_1}+a_2t^{-z_2}$ & $[1.7\times 10^2,*]$ & $\begin{aligned}
            e_\infty = -1.7708874(19),\,a_1 = 0.0884(7),a_2 = 0.1060(3),\\[-4pt] z_1 = 0.2235(5),\,z_2 = 0.445(3)\qquad\qquad\quad\end{aligned}$ & $242.4/248$ \\[5pt]        
        $e$ & $\Tc$ & $e_\infty+a_1t^{-z_1}$ & $[10^4,*]$ & $e_\infty = -1.7022747(19),\,a_1 = 0.1325(12),z_1 = 0.3820(9)$ & $152.4/149$ \\[5pt]
        $e$ & $1.4$ & $e_\infty+a_1t^{-z_1}$ & $[10^4,*] $ & $e_\infty = -1.607026(4),\,a_1 = 2.5(9),z_1 = 0.90(4)$ & $91.7/83$ \\\hline 
        $\xi_{12}$ & $0.7$ & $a_0+a_1\changes{t^z}+a_2\log(t)$ & $[1.4\times 10^2,*]$ & $a_0 = -2.71(3),\,a_1 = 4.10(3),a_2 = -0.1179(14),z = 0.05930(17)$ & $197.6/235$ \\[5pt] 
        $\xi_{12}$ & $\Tc$ & $a_0+a_1\changes{t^z}$ & $[10^3,10^9]$ & $a_0 = 0.159(8),\,a_1 = 1.158(4),z = 0.1543(2)$ & $126.4/157$ \\[5pt] 
              $\xi_{12}$ & $1.4$ & $C\left( 1- \sum_{i=0}^7 a_i\e^{-tb^i/\tau}\right)$ & $[10^3,*]$ & $\begin{aligned}C = 9.29(2),\,\tau = 8(3)\times 10^6,b = 4.0(4),\qquad\qquad\\[-4pt]a_0 = 0.007(4),\,a_1=0,\,a_2 = 0.11(3),a_3 = 0.166(17),\quad \\[-4pt]a_4 = 0.136(9),\,a_5 = 0.123(9),\,a_6 = 0.095(7),a_7 = 0.069(8)\end{aligned}$ & $49.4/104$ 
              \\
              \hline
\end{tabular}
}
\end{table*}
Ideally speaking, we would hope $k^*$ to be close to $\gamma$, because
in a single step of the microcanonical simulated annealing algorithm,
every spin is updated $\gamma$ times (to be compared with just one
spin update per canonical timestep). A direct inspection of
Table~\ref{tab:k*} reveals that only the variants of the algorithm
with walkers get close to this ideal. When one works with daemons
rather than walkers, $k^*$ turns out to be significantly smaller than
$\gamma$. Yet, as we shall discuss in the Conclusions, in order to
choose the most convenient algorithmic variant, one should also take
into account the computational cost of the different parts of the
algorithm (see \ref{appendix:CUDA}).

To better ascertain the quality of the time rescaling, we show in the panels {\bf d} and {\bf e} of Figs.~\ref{fig:T0.7},~\ref{fig:T1.1}, and~\ref{fig:T1.4} the differences
\begin{eqnarray}\label{eq:delta1}
\changes{\delta_e(t;k)}&=&e^\text{MicSA}(t)- F^e_\text{can}(k\,t,T)\,,\\\label{eq:delta2}
\changes{\delta_\xi(t;k)}&=&\xi_{12}^\text{MicSA}(t)- F^\xi_\text{can}(k\,t,T)\,,
\end{eqnarray}
as computed for $k=k^*$. These figures show that, when looked at with enough accuracy, $e^\text{MicSA}(t)$ is \emph{not} a smooth function of time (this is most clear for $T\leq T_\mathrm{c}$). Indeed, these curves have abrupt (but tiny) changes every time the walkers get refreshed. This is why we have not attempted to get the rescaling factor from the energy. These abrupt changes, as well as the difference $\delta_e(t)$ itself, strongly decrease as the time $t$ increases.

\section{Conclusions}
\label{sect:conclusions}
We have proposed a new Monte Carlo algorithm, microcanonical simulated annealing (MicSA), that is both fully adapted to massively parallel computations and extremely frugal in its use of random numbers. The method is fully general and has a certain number of adaptable features. Chief among these are the number $\gamma$ of daemons/walkers per elementary degree of freedom (spins in our case) and the choice between using daemons or walkers. Walkers have the advantage of avoiding local conservation laws and their resulting slower dynamics, but one can decide whether to use daemons or walkers, and how many, on the view of the application at hand and the hardware platform that will be used.

The method performs fully satisfactorily in the demanding setting of the three-dimensional Ising spin glass, where we have shown that it can reproduce the results of a standard (random-number-greedy) Metropolis algorithm both in and out of equilibrium. In particular, we have shown that the out-of-equilibrium dynamics derived from MicSA can be rescaled onto the one from traditional Metropolis by a simple change in the time unit. The fact that, in the case of walkers, the rescaling factor is nearly equal to the ratio of the number of spin updates in a single timestep for both methods (which coincides with $\gamma$) tells us that a single step of MicSA is worth $\gamma$ Metropolis sweeps, without requiring a single random number. 

Our CUDA implementation demonstrates the parallel-friendly nature of
the algorithm. \changes{The code, see \ref{appendix:CUDA}, is publicly
available and may run both
standard Metropolis and MicSA. It achieves a 
performance that could previously only be provided by the custom-built Janus~II
 computer.\footnote{In the context of quantum spin
glasses, performance competitive with Janus II has already been 
obtained on GPU~\cite{bernaschi:24}.} 
Table~\ref{tab:perf} shows the time required for the execution of the two main kernels of the standard Metropolis algorithm and the MicSA algorithm  on a lattice with $160^3$ spins and 128 replicas using a single walker. Using more walkers, the spin-update time increases linearly as expected (data not reported). 

\begin{table}[t]
  \caption{Metropolis and MicSA timings for a full sweep of 128 replicas of a lattice with $160^3$ spins. The test was executed on an Nvidia H100 GPU. The total time of the Metropolis sweep is 1.9 times longer than that of MicSA.
    \label{tab:perf}}
     \centering
     \footnotesize
  \begin{tabular}{ c c r }
  \hline
{Method} & {Kernel} & Time (ns) \\
\hline
 Metropolis  & random-number generation  & 301\,363 \\
 & spin update  &  113\,510  \\
 \hline
 MicSA  & spin update  & 121\,554  \\
 & walking procedure  & 97\,378\\
\hline
\end{tabular}
\end{table}
By adding the timings for the two kernels of the new algorithm and taking into account the total number of spins, we see that the spin-update time is $\approx 0.4$ picoseconds.
This value is much lower than the 3.17
picosecond per spin-update obtained on a single Janus~II FPGA. However, the above two numbers
should be compared with care for two reasons.} First, the
computing power on the GPU is spread over 128 replicas, while on
Janus~II updates are concentrated on only two replicas. Hence, for a
high-accuracy simulation spanning a limited time range (say up to
$t\sim 10^9$), the GPU is faster. If, however, one needs to reach
times $t\sim10^{11}$ or larger (which is not
uncommon~\cite{dahlberg:25}), Janus~II still fares better (the GPU
might make more spin updates per second by running many replicas, but
would need a wall-clock time of many months to finish each
simulation). Second, we are, somewhat unfairly, comparing 2025
high-end GPUs with the 2013 Virtex~7 FPGAs that Janus~II is made of.

Since MicSA is particularly suitable for high-end FPGAs or even dedicated hardware (based on an ASIC, for instance), we expect that users interested in extreme-scale Monte Carlo simulations will find a significant advantage in our proposal. 

The new algorithm raises some intriguing questions, some of which are more theoretical. Let us mention a few of them.

The performance of MicSA for system sizes $L$ small enough to be brought to thermal equilibrium also in the spin-glass phase, $T<T_\mathrm{c}$, needs to be ascertained. In particular, MicSA needs to be combined with Parallel Tempering~\cite{hukushima:96}. We are already testing this algorithm combination for small systems, see \ref{appendix:PTgaussian} and Ref.~\cite{chilin:26}.

The dynamical properties of MicSA in small systems are likely to be quite peculiar. If the daemon/walkers are refreshed at a very low rate, the system may have enough time to equilibrate in the time lapse between the refreshing steps. Since the equilibrium values of the different observables differ in the canonical and the microcanonical ensemble, it is not obvious that it will always be possible to map MicSA dynamics onto canonical dynamics for small systems. Of course, in the long run, canonical equilibrium values will be recovered by MicSA because the daemons/walkers' refresh will vary the total energy.

The performance of MicSA should be investigated in problems where the elementary excitation (which is a spin flip in our case) will cause a continuously distributed energy change. The simplest example would be, perhaps, a spin glass with Gaussian-distributed couplings. As \ref{appendix:PTgaussian} shows, choosing the daemons/walkers as continuous, exponentially distributed random variables fully solves the problem.

Another interesting issue regards the optimal refreshing rate for the daemons/walkers. As we have already explained, if the refresh occurs at a constant time rate, we expect dynamical rescalability with Metropolis. If we refresh at every step with daemons/walkers of infinite capacity, MicSA \emph{is} the Metropolis algorithm (if the capacity is limited, the resulting algorithm is just an unusual version of Metropolis). Of course, if the refreshing rate is too low, microcanonical equilibration will occur in between the refreshing steps, which would complicate the dynamical analysis. However, refreshing once every (say) 1000 steps would bring to microcanonical equilibrium only a fairly small system (and the cost of generating random numbers would be divided by a factor of 1000 as compared to a traditional simulation).

One may wonder how far the extreme strategy followed here can be pushed (recall that we refreshed the daemons/walkers in an exponentially spaced time schedule, namely at times $t_s=2^s$ for $s=0,1,2,3,...$). Given that in, critical systems with a divergent specific heat (\emph{e.g.}, the three-dimensional ferromagnetic Ising model), the microcanonical dynamics cannot be rescaled to coincide with the canonical one~\cite{zinn-justin:05}, it is conceivable that the minimal daemon/walker 
time schedule allowing dynamic scalability will be less spaced than for the spin glass (perhaps $t_s\propto s^a$, with $a$ some appropriate exponent,  would do the job for the three-dimensional Ising model).

\section*{Data availability}
Our CUDA code and the data required to reproduce the figures
are publicly available at~\cite{github}.

\section*{Acknowledgments}
{We thank Martin Hasenbusch for calling our attention to Refs.~\cite{rummukainen:93,caselle:94,hasenbusch:94,agostini:97} and for kindly explaining to us some of the details in these works.}

This work was partially supported by Ministerio de Ciencia, Innovación y Universidades (Spain) and by the European Regional Development Fund (MCIU/AEI/10.13039/501100011033/FEDER, UE) through grants no. 
 PID2022-136374NB-C21, PID2024-156352NB-I00, PID2024-158623NB-C22 and RED2022-134244-T; by Junta de Extremadura (Spain) through grant no.~GR24022; and by funding from the 2021 first FIS (Fondo Italiano per la Scienza) funding scheme (FIS783 - SMaC - Statistical Mechanics and Complexity) from
Italian MUR (Ministry of University and Research). We acknowledge the use of the CESAR computational resources at the BIFI Institute (University of Zaragoza).
IG, MB and FRT acknowledge the support of the National Center for HPC, Big Data and Quantum Computing, Project CN\_00000013 – CUP E83C22003230001, CUP B93C22000620006 and CUP B83C22002940006, Mission 4 Component 2 Investment 1.4, funded by the European Union – NextGenerationEU.

\appendix

\section{Computation of the best time shift}
\label{ap:time-shift}
\renewcommand{\arraystretch}{2}

As we have explained in Sec.~\ref{sect:tests}, to estimate the best value of $k^*$, we use the coherence length $\xi_{12}$. A similar analysis can be performed with the energy, but, as it is clear from panel \textbf{d)} in Fig~\ref{fig:T0.7}, the energy in the MicSA algorithm does not decrease to its equilibrium value monotonically. Instead, for short times, it is sensitive to the daemons' refreshing process. This behavior makes it more challenging to estimate $k^*$ from these data. Nonetheless, the relation between the coherence length and  time in the spin-glass phase, which can be extended to the critical point, offers a straightforward way to estimate $k^*$.

In particular, at the critical point, we can use the following relation~\cite{janus:18}
\begin{equation}
    \log(t) = a_0+z\log(\xi_{12})+\frac{a_1}{\xi_{12}^\omega}\,
    \label{eq:logt_vs_xi}
\end{equation}
where $\omega$ is the exponent characterizing the first corrections to leading scaling and it takes the value $\omega=1.12(10)$~\cite{janus:13}. 

For $T<\Tc$,  the leading behavior is given by $\omega=\theta$, where $\theta$ is the replicon exponent (we used $\omega=0.35$ in the fits~\cite{janus:18}).

As the reader will notice, it is easy to include in Eq.~\eqref{eq:logt_vs_xi} a time shift $k$ that connects our MicSA data with the canonical curve:
\begin{equation}
    \log(t) = a'_0-\log(k)+z'\log(\xi_{12})+\frac{a'_1}{\xi_{12}^\omega}\,.
    \label{eq:logt_vs_xi_shift}
\end{equation}
In this situation, we propose the following method to extract the best time shift, $k^*$:
\begin{enumerate}
    \item Fit the canonical data to Eq.~\eqref{eq:logt_vs_xi} in order to obtain  parameters $a_0$, $a_1$, and $z$.
    \item With parameters fixed to the values obtained in the previous step ($a'_0=a_0$, $a_1'=a_1$, and $z'=z$), we perform a new fit to the MicSA data to Eq.~\eqref{eq:logt_vs_xi_shift}. 
\end{enumerate}
This process guarantees that the only fitting parameter in step 2 is $k$, obtaining $k^*$. Table~\ref{tab:k*} summarizes the results of $T\leq\Tc$. Fig.~\ref{fig:fit_k*} reproduces this procedure. In the main panel, we report the difference between our MicSA data $\log(t_{\mathrm{MicSA}})$ and the interpolation $f(t;k^*|\Theta)$, see Eq.~\eqref{eq:logt_vs_xi_shift}, with the best time shift $k^*=5.875$ and the parameters $\Theta=(a_0,a_1,z)$ obtained following this process for $\gamma=6$ and $T=\Tc$. The inset shows the non-subtracted data.

This approach is, however, not useful at $T>\Tc$, because Eq.~\eqref{eq:logt_vs_xi} does not hold. In this regime, we interpolate $\xi_{12}(t)$ as a function of time using the following expression
\begin{equation}
    \changes{F^\xi_{\text{can}}(t)} = C\left( 1- \sum_{i=0}^7 a_i\e^{-tb^i/\tau}\right)\,.
    \label{eq:xi_sum}
\end{equation}
In this situation, we can use a similar protocol to extract the value of $k^*$:
\begin{enumerate}
    \item We fit our canonical data to Eq.~\eqref{eq:xi_sum} to obtain the values of the parameters (see the fit result in Table~\ref{tab:fit-info}).
    \item Using the parameters obtained in the previous step, we fit our MicSA data to \changes{$F^\xi_{\text{can}}(kt)$}, where $k$ is the only fitting parameter.
\end{enumerate}
 As in the case of $T\leq\Tc$, this process allows us to obtain directly the best time shift $k^*$, even in the case of $T>\Tc$. Notice that, in this situation, $t$ is the independent variable. Then, in Table~\ref{tab:k*} we reported the fitting range in $t$, instead of $\xi_{12}$. Notice that, although we use a slightly different process to obtain the value of $k^*$ for $T>\Tc$, panels \textbf{d} and \textbf{e} in Fig.~\ref{fig:T1.4} reported a value of $\delta_e$ and $\delta_{\xi_{12}}$ compatible with zero at long times.

\section{The CUDA implementation}\label{appendix:CUDA}
In the following, we provide basic technical information about the CUDA implementation of the MicSA algorithm described in Sec.~\ref{sect:model_and_algorithm} applied to the three-dimensional Ising spin glass. The code implements a two-level parallelism: we use multispin coding to pack 64 spins in 64-bit words (\verb |long long int|) and then CUDA threads for the concurrent update of {\em even} and {\em odd} spins.
Multiple {\em replicas} and {\em samples} can be simulated in a single execution; the actual number of both depends on the resources (mainly memory) available on the GPU (we successfully tested up to 512 replicas). The code also implements the standard Metropolis algorithm using random numbers generated on the GPU with the procedure described in Ref.~\cite{bernaschi:24}.
The refresh of the daemons/walkers is carried out on the CPU asynchronously to the microcanonical update (we recall that the refresh takes place according to an exponential schedule) using CPU threads. The walking procedure and the microcanonical update are implemented in two separate CUDA kernels to increase flexibility. The main reason is that the microcanonical update is carried out separately for each daemon/walker, whereas the walking procedure is completed for all the daemons in a single kernel execution.

The walking procedure takes $\approx44\%$ of the execution time on the GPU. In contrast, random-number generation takes the majority ($\approx 72\%$) of the time in the Metropolis algorithm.  We argue that the higher the peak performance of the platform, the greater the advantage provided by minimizing the use of random numbers. In particular, the design of an \changes{{\em Application Specific Integrated Circuit} (ASIC)} would be greatly simplified under the assumption that generating random numbers is not an issue.

Further details about our code can be obtained by direct inspection of the source available from~\cite{github}.

\section{A Parallel Tempering example (walkers in equilibrium simulations)}\label{appendix:PTgaussian}

\begin{figure*}[bt]
    \centering
    \includegraphics[width=\linewidth]{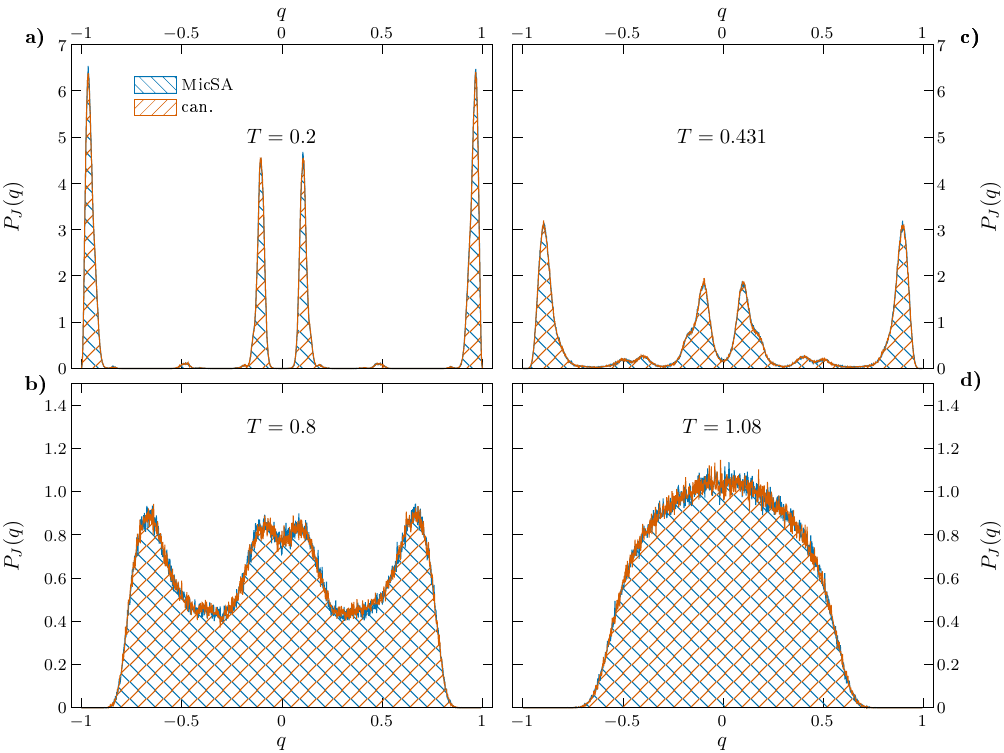}
    \caption{{Overlap distribution of some selected temperatures}. The curves for each method are obtained by simulating 200 times the same sample for 250000 EMCS (half of which have been used for prethermalization, meaning that we keep data only from the second half of the simulation) and later averaging over the 200 curves obtained. The simulation parameters are $L=10$, $N_R=2$, $N_\beta = 40$, temperatures uniformly distributed between $T_{\min}=1/\beta_{\max} = 0.2$ and $T_{\max}=1/\beta_{\min} = 2$.}
    \label{fig:p_q_PTH}
\end{figure*}

In this section, we provide an example of the possible use of the
walkers for a simulation at equilibrium, which radically
differs from the off-equilibrium dynamics studied in the main
text. Bringing a spin-glass sample to equilibrium is possible only if the linear size of the system is tiny (particularly at low temperatures). Furthermore, special simulation methods, such as
Parallel Tempering~\cite{hukushima:96}, are mandatory. The particular
problem that we shall investigate here is a simple variant of
Eq.~\eqref{eq:Hamiltonian} in which the couplings are drawn from a
Gaussian distribution $J \sim \mathcal{N}(0,1)$. The rationale for
choosing Gaussian-distributed couplings is that we wish to test our
methods in a demanding situation, namely at the very low temperature 
$T=0.2\approx 0.21 T_\mathrm{c}$ ($T_\mathrm{c}$ is the critical
temperature). Indeed, continuously distributed couplings are usually
preferable for very-low-temperature studies.

Specifically, we shall study in detail a single sample of a 3D system with
lattice linear size $L=10$.  For this sample, the walkers method reproduces the equilibrium values of the energy per spin [\emph{i.e.}, the spin Hamiltonian in Eq.~\eqref{eq:Hamiltonian} divided by the number of spins $N$] and of the overlap between the instantaneous value of two replicas \begin{equation}
    q = \frac{1}{N} \sum_{\boldsymbol{x}} s_{\boldsymbol{x}}^{(a)} s_{\boldsymbol{x}}^{(b)}.
\end{equation}
In order to do this, we adapted the walker method to the Parallel
Tempering procedure enhanced with Houdayer moves \cite{houdayer:01},
which has been proven to be (mildly) effective also for the
equilibrium simulation of three-dimensional spin glasses
\cite{zhu:15}, even though our schedule is a bit different
\cite{chilin:26}. We are then running $N_R=2$ replicas of
$N_\beta = 40$ clones with temperatures uniformly distributed between
$T_{\min}=1/\beta_{\max} = 0.2$ and $T_{\max}=1/\beta_{\min} = 2$. 

In this case, our method requires the use of continuous walker variables, distributed as in Eq.~\eqref{eq:canonical2}. Additionally, we do not need to limit the capacity of the walkers as in Sec.~\ref{subsec:limitedcapacity}, but we can take a step back and use the elementary move described in Sec.~\ref{subsec:creutzmove}. 

\begin{table}[t]
  \caption{Measures of  the equilibrium expectation values for $|q|$, $e_\infty$ as extracted from the simulations described in  \ref{appendix:PTgaussian}, at the lowest temperature $T_\text{min}=0.2$. For each of the two algorithms, we carried out 200 independent runs. Errors are straightforwardly computed by considering the fluctuations between runs. 
\label{tab:ptobs}}
\centering
\footnotesize
  \begin{tabular}{lllll}
  \hline
            & $|q|$     & error & $e_\infty$  & error     \\ \hline
Walkers    & 0.743\,8  & 0.012\,9  & $-1.696\,330\,17$ & 0.000\,001\,02  \\
Metropolis & 0.741\,6  & 0.010\,4  & $-1.696\,331\,13$ & 0.000\,000\,89   \\
Difference & 0.002\,19 & 0.016\,7 & $\phantom{-}0.000\,000\,96$   & 0.000\,001\,36\\
\hline
\end{tabular}
\end{table}

For each replica/clone, we simulate a system with one walker per spin ($\gamma=1$) that gets refreshed at each Elementary Monte Carlo Step (EMCS) as described below.

\changes{The EMCS is structured as follows:}
\begin{enumerate}
    \item For each clone and for each replica, perform a fixed-temperature move:
    \begin{enumerate}
        \item Generate the walkers according to Eq.~\eqref{eq:canonical2}.
        \item Alternate a full-lattice sweep with a wandering of the walkers towards the foward-X direction. Repeat 10 times.
    \end{enumerate}
    \item Perform Houdayer's move.
    \item Perform a Parallel Tempering step between the $N_\beta$ clones belonging to the same replica group.
\end{enumerate}

Fig.~\ref{fig:p_q_PTH} qualitatively shows how the probability
distribution of $q$ computed in this manner reproduces the results
of using a canonical Metropolis update instead of point 1 of the above algorithm.

As a quantitative test, we have also evaluated the average
energy per spin and the average overlap (in absolute value) yielded by
both simulations. The obtained values are summarized in
Table~\ref{tab:ptobs}, which displays good agreement between the two
algorithms. Furthermore, from the ratio between the errors we can
estimate the ratio of the \changes{integrated autocorrelation} time for both
algorithms~\cite{sokal:97}.  We obtain for $|q|$ the
ratio $\tau_\text{walkers}/\tau_\text{Metropolis} \simeq 1.7$, meaning that, in
this setting, one EMCS with the walkers is roughly equivalent to $0.59$
steps with the usual Metropolis.  We can understand which of the two
ways is more effective only by measuring also the CPU time for one
EMCS, which depends not only on hardware, but also on
parameters of the simulation such as the linear size $L$ or the number
of temperatures $N_\beta$ of the Parallel Tempering.\footnote{For
  instance, our partially multi-spin coded program runs about twice as
  fast with walkers as with Metropolis when $L=18$ on an {\tt AMD
    EPYC 7763 64-Core Processor}; the computational advantage of the
  walker algorithm seems to grow with $L$.}

\end{document}